\documentclass{article}
\usepackage{spconf,amsmath,graphicx,hyperref}
\usepackage[numbers]{natbib}
\usepackage{enumitem}
\usepackage{pgfplots}
\pgfplotsset{compat=1.18}
\usepackage{xcolor}
\usepackage{booktabs,multirow,colortbl,xcolor}
\usepackage{pgfplots}
\usepackage{booktabs}
\usepackage{pgfplotstable}
\usepackage{titlesec}
\definecolor{better}{RGB}{0,150,0} 
\definecolor{worse}{RGB}{200,0,0}  
\usepackage{tablefootnote}

\usepackage{array}      

\usepackage{multirow}
\usepackage{makecell} 

\title{Can Large Audio Language Models Understand Child Stuttering Speech? Speech Summarization, and Source separation}

\name{
  Chibuzor Okocha$^{1}$, 
  Maya Bakri$^{2}$, and
  Christan Grant$^{1}$
}
\address{
  $^{1}$Department of Computer Science, University of Florida, Gainesville, FL, USA\\
  $^{2}$Department of Computer Science, Lebanese American University
}

\begin{document}
%
\maketitle
\begin{abstract}
Child speech differs from adult speech in acoustics, prosody, and language development, and disfluencies (repetitions, prolongations, blocks) further challenge Automatic Speech Recognition (ASR) and downstream Natural Language Processing (NLP). Recent large audio-language models (LALMs) demonstrate strong cross-modal audio understanding; however, their behavior in disfluent child speech remains underexplored. We evaluate several state-of-the-art LALMs in two settings: an interview (mixed speakers) and a reading task (single child). The tasks are (i) single-channel source separation to isolate the child and (ii) child-only summarization that preserves clinically relevant disfluencies and avoids adult-speech leakage.

Evaluation combines Large Language Model (LLM) as a judge, human expert ratings, and BERTScore (F1), and we report agreement between models and between models and humans to assess reliability.
Our findings delineate the conditions under which LALMs produce faithful child-only summaries from mixed audio and where they fail, offering practical guidance for clinical and educational deployments. We provide prompts and evaluation scripts to support replication.
\end{abstract}
\begin{keywords}
Large Audio-Language Model, LLM, Source Separation, Summarization, Child Speech, Stutter
\end{keywords}

\begin{figure}[!t]
    \centering
    \includegraphics[width=\linewidth]{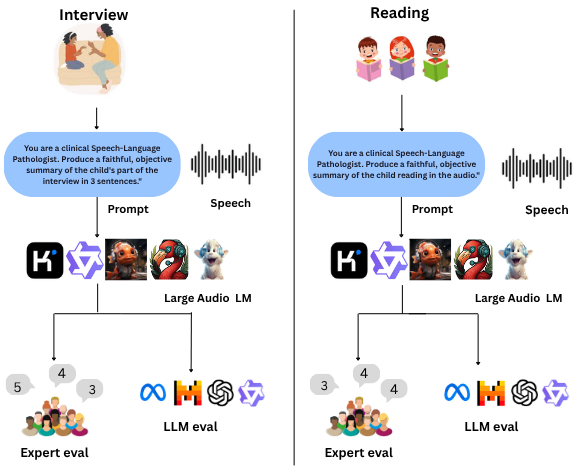}
    \caption{Overview of the evaluation, including the LALMs used for summarization and LLMs and Human expert evaluations.}
    \label{fig:evaluation}
\end{figure}

\section{Introduction}
\label{sec:intro}

Recent advances in ASR and LALMs have opened up new opportunities for tools that can enhance pediatric health and education services \cite{yin_can_2025, sanni_afrispeech-dialog_2025, okocha2025afrivox}. However, robust modeling of children’s speech remains challenging. Relative to adults, children show higher and more variable fundamental frequency \cite{gerosa2007childASR}, evolving articulation and smaller or shifted vowel spaces \cite{vorperian_vowel_2007}, as well as slower and less stable speaking rates \cite{lee_acoustics_1999}. These developmental differences, combined with marked inter-child variability by age, context, and clinical presentation \citep{potamianos2003children, gerosa2007childASR}, create persistent domain shifts that degrade adult-optimized ASR and downstream applications.

At the same time, LALMs have begun to demonstrate unified audio understanding, instruction following over long audio contexts, and speech-centric reasoning. Models such as Qwen2-Audio \cite{qwen2audio2024}, SALMONN \cite{salmonn2023}, and Audio Flamingo~3 \cite{af3_2025} accept raw audio and natural-language instructions, exhibiting emergent abilities for multi-speaker comprehension and long-context summarization \citep{sakshi_mmau_2024, yang_air-bench_2024}. Building on these developments, we consider a real and consequential use case: producing \emph{child-only} summaries from multi-speaker child–adult interviews, where faithfulness to the child’s content and avoidance of adult-speech leakage are critical.

Within clinical and educational settings, stuttering and other fluency disorders introduce additional phenomena repetitions, prolongations, blocks, and secondary behaviors—that stress both recognition and summarization \cite{yairi_early_1999}. Community resources, including UCLASS and FluencyBank (with recent timestamped updates), make possible both descriptive and predictive work on disfluency while underscoring the need for speaker-aware, content-faithful outputs that neither erase disfluency nor hallucinate information not present in the child’s speech \citep{howell2009uclass, fluencybank2024}. Because our downstream objective is a \emph{child-only} summary from mixed audio, the modeling challenge is not only “what was said” but also “\emph{who} said it.” Imperfect diarization or source separation can cause \emph{adult-speech leakage} \cite{okocha_domain-aware_2025}, in which interviewer prompts are paraphrased or attributed to the child.

We propose and evaluate \emph{speaker-purity-aware} child-only summarization from mixed child–adult interviews. To our knowledge, this is the \emph{first} study to explore LALMs for child speech analysis in the presence of stuttering/fluency disorders. Concretely, we:
\begin{itemize}[nosep,leftmargin=*]
\item Formalize child-only summarization with explicit \textbf{speaker purity} constraints and instantiate two reproducible pipelines: {text-first} , {audio-first} (LALMs with child-focused prompting).
\item Introduce an evaluation framework that complements “LLM-as-judge” with expert reference summaries and task-grounded metrics: \textbf{content-unit coverage}, \textbf{speaker leakage/purity}, and \textbf{faithfulness checks}, alongside human ratings with inter-rater reliability.
\item Benchmark multiple audio LLMs and text baselines; and provide targeted error analyses (leakage, omission, hallucination) that identify conditions where LALMs produce the most faithful child-only summaries \cite{shahnawazuddin20_interspeech, okocha_domain-aware_2025}.
\end{itemize}

Our results highlight practical choices for deploying LALM-based tools in pediatric contexts, including when to prefer text-first versus audio-first approaches, the importance of diarization/separation quality for leakage control, and which prompt designs best preserve disfluency while remaining concise, accurate, and clinically useful.

\section{Background and Related Work}
\label{sec:format}

Traditional ASR systems are designed to convert speech into text \cite{ren2019almost}. Once converted, the text can then be used for other NLP tasks, such as summarization, question answering, and sentiment analysis \cite{patil2022state}. Over time, ASR systems have become increasingly advanced. Modern models, such as Transformer-Transducer \cite{zhang2020transformer}, Whisper \cite{radford2023robust}, and Conformer \cite{gulati2020conformer}, achieve near-human accuracy on adult speech but show substantial performance degradation on spontaneous or pediatric speech due to domain mismatch \cite{basak2023challenges}.

Recently, LALMs have emerged, combining powerful speech encoders with large language model decoders to understand and generate language directly from raw audio \cite{bhati2024state}. These models extend far beyond simple transcription: they learn to interpret nuanced acoustic cues—such as tone, pitch, and rhythm—and use these signals to reason, summarize, and generate coherent, context-aware responses \cite{peng2024survey,yang2025towards}. As highlighted in surveys by Peng et al. \cite{peng2024survey} and Yang et al. \cite{yang2025towards}, this development marks a shift from traditional “speech-to-text” systems toward “speech-to-meaning” models that can interpret not only the words spoken but also the surrounding context, emotion, and intent.

Despite these advances, both ASR systems and LALMs continue to underperform on child speech, largely due to domain mismatch in adult-trained data \cite{bhardwaj2022automatic,yeung2018difficulties}. Recent efforts such as CHSER \cite{shankar2025chser} and DRAF \cite{fan2022towards} have improved transcription through data augmentation and self-supervised adaptation, yet they remain focused on recognition accuracy rather than comprehensive multimodal understanding \cite{peng2024survey,yang2025towards}.
\setlength{\parskip}{0pt}

Comprehensive understanding extends beyond recognition, encompassing source separation, summarization, and audio captioning \cite{peng2024survey,yang2025towards}. Source separation distinguishes a child’s voice from overlapping adults or background noise \cite{luo2019conv}, while summarization condenses spoken content into concise, faithful representations. Audio captioning further describes the surrounding acoustic scene and speaker context \cite{huang2024audiogpt}. By integrating these abilities through shared acoustic–semantic embeddings, LALMs achieve a deeper form of multimodal understanding \cite{peng2024survey,yang2025towards}.
\setlength{\parskip}{0pt}

While Large Audio–Language Models (LALMs) have shown strong generalization across a range of domains \cite{peng2024survey,huang2024audiogpt}, their behavior on natural, disfluent child speech remains largely unexplored \cite{yeung2018difficulties}. Existing benchmarks rarely test whether these models can effectively (a) separate a child’s voice from overlapping speakers or (b) generate accurate, child-focused summaries \cite{peng2024survey,yang2025towards}. Addressing this gap is essential for developing reliable and inclusive systems that can understand children’s speech in real-world clinical and educational contexts.


\section{Methods}
\label{sec:methods}

\subsection{Tasks and Settings}
We study two child–speech settings: (a) \textbf{Reading}, where a single child speaker reads aloud, and (b) \textbf{Interview}, a dialog between a child and an adult. We evaluate two capabilities of Large Audio–Language Models (LALMs): (1) \emph{single-channel source separation} in Interview audio, to isolate the child’s speech from mixed speakers; and (2) \emph{child-only summarization}, to produce concise, clinically useful summaries without adult content and without over-normalizing disfluencies (repetitions, prolongations, blocks).

\subsection{Dataset: FluencyBank English Voices}
\label{subsec:data}

We use the FluencyBank English Voices--CWS corpus , a curated collection of video recordings and aligned transcripts of children who stutter (CWS) collected in the United States \cite{fluencybank2024}. The corpus was assembled by Nan Bernstein Ratner, with recruitment and logistical support from FRIENDS (a national organization for children and families who stutter) and the National Stuttering Association, and recordings were obtained at their annual meetings.

\paragraph{Participants and design.}
The dataset comprises \textbf{22} child participants (CWS). Each participant completed two tasks: (i) a semi-structured \textit{interview} (\texttt{int}) and (ii) a \textit{reading} task (\texttt{rdg}). The interview follows a fixed six-question protocol eliciting experiences and perspectives about stuttering:\footnote{Prompts summarized from the corpus page: activities at the FRIENDS meeting; family reasons for attending; peer interactions; talking about one’s speech; what was learned about stuttering; speech therapy experiences and advice for future clinicians.} 
The reading task uses grade-appropriate passages from the Stuttering Severity Instrument--4 (SSI--4) under license from the publisher.\footnote{Per corpus policy and publisher license, we do not reproduce SSI--4 passage text in this paper.}

\subsection{Models}
\label{subsec:models}

We evaluate two families of systems: (i) \emph{audio-first} LALMs that operate directly on waveform inputs (\textbf{Qwen2-Audio}, \textbf{SALMONN}, \textbf{GAMA}, \textbf{Audio Flamingo--3}, \textbf{Kimi Audio}), and (ii) a \emph{text-first} baseline that transcribes and separates before summarizing (Whisper Large $\rightarrow$ pyannote.audio $\rightarrow$ Llama-3.2). For evaluation, we additionally use \emph{LLM-as-Judge} models (\textbf{Qwen2-7B}, \textbf{Mistral-7B}, \textbf{Llama-3.2}) under a shared rubric. A consolidated summary of modality and concise Type/Architecture/Params for each model appears in Table~\ref{tab:model_overview}.

\begin{table*}[t]
\centering
\caption{Models grouped by role. References appear inline with model names. ``Type / Architecture / Params''}
\label{tab:model_overview}
\footnotesize
\setlength{\tabcolsep}{6pt}
\renewcommand{\arraystretch}{1.15}
\begin{tabular}{@{}l l l >{\raggedright\arraybackslash}p{9.8cm}@{}}
\toprule
\textbf{Group} & \textbf{Model (with ref)} & \textbf{Modality} & \textbf{Type / Architecture / Params (short description)} \\
\midrule
\multirow{5}{*}{Audio-first LALMs}
& Qwen2-Audio~\cite{qwen2audio2024} & Audio+Text
& Instruction-following audio–language model; audio encoder + Qwen2 LLM backbone with cross-modal attention for long-context comprehension and summarization. Public checkpoints in multiple sizes (specify the one used). \\

& SALMONN~\cite{salmonn2023} & Audio+Text
& Multi-task audio–language system; pretrained speech encoder fused to an LLM via adapters for open-ended audio QA and summaries; supports instruction prompts. (Params vary by LLM backbone; specify your variant.) \\

& GAMA \tablefootnote{\tiny\url{https://huggingface.co/papers/2406.11768}} & Audio+Text
& General audio–language model for open-ended audio understanding and summarization; encoder–decoder fusion with instruction tuning. (Public size varies; insert exact params for your checkpoint if available.) \\

& Audio Flamingo--3~\cite{af3_2025} & Audio+Text
& Flamingo-style cross-attention between audio features and language backbone for long-context audio reasoning/summarization. (Report the backbone and parameter count used.) \\

& Kimi Audio \tablefootnote{\tiny\url{https://huggingface.co/moonshotai/Kimi-Audio-7B-Instruct}} & Audio+Text
& Production audio assistant; multi-speaker audio comprehension and summarization; black-box API / model card with limited disclosed params. (If known, add backbone + size.) \\
\midrule
\multirow{3}{*}{Text-first baseline}
& Whisper Large \tablefootnote{\tiny\url{https://huggingface.co/openai/whisper-large}} & Audio$\rightarrow$Text
& ASR with word-level timestamps; encoder–decoder Transformer trained on weakly supervised multilingual data (reported $\sim$1.5B params for Large). \\

& pyannote.audio \tablefootnote{\tiny\url{https://huggingface.co/pyannote/speaker-diarization-3.1}} & Audio
& Neural diarization toolkit; VAD + embedding extractor + clustering; commonly x-vector or ECAPA-TDNN style embeddings. Used here to isolate child turns. (No single param count—pipeline of modules.) \\

& Llama-3.2 (Instruct) \tablefootnote{\tiny\url{https://huggingface.co/meta-llama/Llama-3.2-3B-Instruct}} & Text
& Instruction-tuned text LLM used for summarization over concatenated child turns. (Insert your exact family/size, e.g., Llama-3 Instruct 8B/70B.) \\
\midrule
\multirow{3}{*}{LLM-as-Judge}
& Qwen2-7B \tablefootnote{\tiny \url{https://huggingface.co/Qwen/Qwen2-7B-Instruct}} & Text
& Open LLM used only as a judgment model; rubric-prompted for faithfulness, coverage, coherence, usefulness. (~7B params.) \\

& Mistral-7B \tablefootnote{\tiny\url{https://huggingface.co/mistralai/Mistral-7B-v0.1}} & Text
& Open LLM judge; dense Transformer with sliding-window attention; rubric-prompted scoring. (~7B params.) \\

& Llama-3.2 (Instruct)  & Text
& Open LLM judge; instruction-tuned; rubric-prompted scoring. (Insert exact size used, e.g., 8B/70B.) \\
\bottomrule
\end{tabular}
\end{table*}

\subsection{Prompts}
\label{subsec:prompts}

We standardize prompts across systems and report them verbatim. Decoding is fixed unless otherwise noted: temperature$=0.2$, top\_p$=0.95$, max tokens$=512$. The prompting techniques used were zeroshot for the interview and few shot for the LLM as a judge.

\paragraph{Interview and Reading summarization.}
\begin{quote}\small
\textbf{System role:} You are a clinical Speech--Language Pathologist.\\
\textbf{Task:} Produce a faithful, objective 3--sentence summary of the child's \emph{reading} performance (content and difficulties) from the transcript/audio.\\
\textbf{Schema:} \texttt{\{}"summary": string\texttt{\}}.\\
\textbf{Guidelines:} Keep the summary concise, faithful, and specific to the child's speech. (Adult-filtering is typically unnecessary for reading items.)
\end{quote}

\paragraph{LLM-as-Judge (evaluation).}
\begin{quote}\small
\textbf{System role:} You are an expert Speech--Language Pathologist and evaluation judge for child-speech summarization.\\
\textbf{Instruction:} Evaluate the \textsc{Model Summary} against the \textsc{Reference Summary} created by a baseline text LLM. Rate each criterion from 1 (very poor) to 5 (excellent): (1) Overall quality; (2) Fluency/Coherence; (3) Faithfulness/Factuality; (4) Coverage of child’s main points; (5) Speaker purity (child-only); (6) Usefulness (for SLPs/parents/researchers).\\
\textbf{Inputs:}\\
\textsc{Reference Summary:} \{\texttt{reference}\}\\
\textsc{Model Summary:} \{\texttt{summary}\}\\
\textbf{Output (strict JSON):}\\
\texttt{\{}"score\_overall": int, "score\_fluency": int, "score\_faithfulness": int, "score\_coverage": int, "score\_purity": int, "score\_usefulness": int, "rationale": string\texttt{\}}
\end{quote}

\subsection{Evaluation}
\label{subsec:evaluation}

\paragraph{Automatic metrics.}
We assess summary quality against baseline summary using \textbf{BERTScore (F1)} and \textbf{ROUGE}.\footnote{We report ROUGE--1/2/Lsum as F1, and include precision/recall for transparency.}
Concretely, for each system and item we compute:
(i) \textbf{ROUGE--1/2/Lsum} (\texttt{rouge1\_f1}, \texttt{rouge2\_f1}, \texttt{rougeLsum\_f1}) alongside \texttt{precision}, \texttt{recall}, and \texttt{f1}; and
(ii) \textbf{BERTScore--F1} as a semantic adequacy proxy.
Uncertainty is estimated via a nonparametric bootstrap over items ($1{,}000$ resamples); we report \texttt{f1\_ci\_low} and \texttt{f1\_ci\_high} as 95\% confidence intervals.

\paragraph{Human ratings and LLM-as-Judge.}
Two human experts (speech/fluency) rate each summary on a shared rubric (faithfulness, coverage, coherence, usefulness, and speaker purity).
In parallel, \textit{Qwen2--7B}, \textit{Mistral--7B}, and \textit{Llama--3.2} serve as LLM judges with the same rubric and anonymized, randomized system outputs.

\textbf{Reliability and agreement.}
We quantify within-group reliability and cross-group agreement as follows:
\begin{itemize}[leftmargin=*]
  \item \textbf{LLM vs LLM:} Pairwise \textbf{correlations} between judge models using \texttt{Pearson\_r} (with $p$-values) and \texttt{Spearman\_r}, computed per score dimension (\texttt{overall}, \texttt{fluency}, \texttt{faithfulness}, \texttt{coverage}, \texttt{purity}, \texttt{usefulness}) and per task (\texttt{task} $\in$ \{\textsc{Interview}, \textsc{Reading}\}). We also report \texttt{Cohen\_kappa} after discretization when used.
  \item \textbf{LLM vs Human:} Concordance between each judge model and the human mean via \texttt{Pearson\_r}/\texttt{Spearman\_r} and (when preferences are collected) accuracy and $\kappa$ on pairwise comparisons.
\end{itemize}
For all agreement statistics we include the sample size (\texttt{n\_samples}) and 95\% CIs via bootstrap. Task-level results are reported separately for \textsc{Interview} and \textsc{Reading} and then macro-averaged across tasks in the main Table~\ref{tab:judge_scores_reading} and Table~\ref{tab:interview_judge_full_nocov}.

\subsection{Reproducibility}
We release prompts, scoring scripts (BERTScore; ROUGE), judge wrappers, and bootstrap utilities. Configuration files specify temperatures, maximum tokens, and random seeds; compute details (GPU type, batch sizes, runtime per hour of audio) are documented in the github repo.

\begin{table*}[htbp]
\centering
\caption{Average Scores (1--5) from Each LLM-as-Judge for \textsc{Reading} Summaries. 
Best per column in \textbf{bold}.}
\label{tab:judge_scores_reading}
\resizebox{\linewidth}{!}{
\begin{tabular}{lccccc ccccc ccccc}
\toprule
\multirow{2}{*}{\textbf{Model}} &
\multicolumn{5}{c}{\textbf{Llama-3.2-8B-Instruct}} &
\multicolumn{5}{c}{\textbf{Mistral-7B-Instruct-v0.3}} &
\multicolumn{5}{c}{\textbf{Qwen2-7B-Instruct}} \\
\cmidrule(lr){2-6}\cmidrule(lr){7-11}\cmidrule(lr){12-16}
 & \textbf{Overall} & \textbf{Fluency} & \textbf{Faithfulness} & \textbf{Coverage} & \textbf{Usefulness}
 & \textbf{Overall} & \textbf{Fluency} & \textbf{Faithfulness} & \textbf{Coverage} & \textbf{Usefulness}
 & \textbf{Overall} & \textbf{Fluency} & \textbf{Faithfulness} & \textbf{Coverage} & \textbf{Usefulness} \\
\midrule
\rowcolor{blue!5}
AF35 & 2.74 & \textbf{3.74} & 2.21 & 2.16 & 2.63 & 
\textbf{3.11} & \textbf{4.16} & \textbf{2.58} & \textbf{2.26} & 2.11 &
\textbf{3.84} & \textbf{4.68} & \textbf{3.26} & \textbf{3.79} & \textbf{4.26} \\
\rowcolor{blue!2}
Salmon & 2.73 & 3.46 & 2.05 & \textbf{2.36} & \textbf{2.73} &
2.77 & 3.59 & 2.50 & 2.18 & \textbf{2.27} &
3.50 & 4.18 & 2.77 & 3.32 & 3.64 \\
Qwen & 2.32 & 2.91 & 1.77 & 1.73 & 1.96 &
2.23 & 3.23 & 2.05 & 1.55 & 1.68 &
3.50 & 4.36 & 2.68 & 3.27 & 3.55 \\
GAMA & 1.68 & 1.91 & 1.23 & 1.18 & 1.50 &
1.09 & 1.05 & 1.00 & 1.00 & 1.00 &
3.05 & 3.91 & 1.96 & 2.86 & 2.96 \\
Kimi & \textbf{2.91} & 3.41 & \textbf{2.46} & 2.09 & 2.50 &
2.27 & 3.18 & 2.05 & 1.36 & 1.50 &
3.73 & 4.46 & 3.23 & 3.41 & 3.82 \\
\bottomrule
\end{tabular}}
\end{table*}

\begin{table*}[htbp]
\centering
\caption{Average Scores (1--5) from Each LLM-as-Judge for \textsc{Interview} Summaries. 
Best per column in \textbf{bold}.}
\label{tab:interview_judge_full_nocov}
\resizebox{\linewidth}{!}{
\begin{tabular}{l ccccc ccccc ccccc}
\toprule
\multirow{2}{*}{\textbf{Model}} &
\multicolumn{5}{c}{\textbf{Llama-3.2-8B-Instruct}} &
\multicolumn{5}{c}{\textbf{Mistral-7B-Instruct-v0.3}} &
\multicolumn{5}{c}{\textbf{Qwen2-7B-Instruct}} \\
\cmidrule(lr){2-6}\cmidrule(lr){7-11}\cmidrule(lr){12-16}
& \textbf{Overall} & \textbf{Fluency} & \textbf{Faithfulness} & \textbf{Purity} & \textbf{Usefulness}
& \textbf{Overall} & \textbf{Fluency} & \textbf{Faithfulness} & \textbf{Purity} & \textbf{Usefulness}
& \textbf{Overall} & \textbf{Fluency} & \textbf{Faithfulness} & \textbf{Purity} & \textbf{Usefulness} \\
\midrule
\rowcolor{blue!5}
AF3 & \textbf{3.50} & 3.92 & \textbf{2.88} & 4.67 & \textbf{3.42} &
\textbf{3.00} & \textbf{4.50} & \textbf{2.92} & \textbf{4.88} & \textbf{3.29} &
3.79 & 4.42 & 3.42 & 4.63 & \textbf{3.58} \\
\rowcolor{blue!2}
SALMONN & 2.23 & 3.08 & 1.46 & 4.08 & 2.12 &
1.96 & 3.50 & 1.77 & 4.65 & 2.08 &
3.23 & 4.08 & 2.38 & 4.42 & 2.82 \\
Qwen & 2.31 & 3.23 & 1.42 & \textbf{4.69} & 2.19 &
2.08 & 4.54 & 1.69 & 4.71 & 2.15 &
3.23 & 4.12 & 2.42 & 4.53 & 2.44 \\
GAMA & 1.88 & 2.19 & 1.04 & 4.10 & 1.88 &
1.58 & 1.69 & 1.08 & 4.22 & 1.74 &
2.85 & 3.65 & 1.88 & 4.00 & 1.92 \\
Kimi & 2.96 & \textbf{3.62} & 2.21 & 4.33 & 3.00 &
2.79 & 4.50 & 2.58 & 4.60 & 3.08 &
\textbf{3.63} & \textbf{4.21} & \textbf{3.21} & \textbf{4.52} & 3.20 \\
\bottomrule
\end{tabular}}
\end{table*}

\begin{table}[htbp]
\centering
\caption{Average ($\pm$SD) Scores Across LLM Judges (Llama-3.2, Mistral, Qwen) for \textsc{Reading} Summaries. 
Best per column in \textbf{bold}.}
\label{tab:reading_llm_judges}
\resizebox{\linewidth}{!}{
\begin{tabular}{lccccc}
\toprule
\textbf{Model} & \textbf{Overall} & \textbf{Fluency} & \textbf{Faithfulness} & \textbf{Coverage} & \textbf{Usefulness} \\
\midrule
\rowcolor{blue!5}
AF35   & \textbf{3.23 (±0.56)} & \textbf{4.19 (±0.48)} & \textbf{2.68 (±0.43)} & \textbf{2.74 (±0.69)} & \textbf{3.00 (±0.91)} \\
\rowcolor{blue!2}
SALMON & 3.00 (±0.40) & 3.74 (±0.30) & 2.44 (±0.31) & 2.62 (±0.49) & 2.88 (±0.61) \\
Qwen   & 2.68 (±0.65) & 3.50 (±0.59) & 2.17 (±0.49) & 2.18 (±0.76) & 2.40 (±0.85) \\
GAMA   & 1.94 (±0.83) & 2.29 (±1.21) & 1.40 (±0.55) & 1.68 (±0.94) & 1.82 (±1.06) \\
\rowcolor{blue!5}
Kimi   & 2.97 (±0.73) & 3.68 (±0.66) & 2.58 (±0.65) & 2.29 (±0.84) & 2.61 (±0.99) \\
\bottomrule
\end{tabular}}
\end{table}

\begin{table}[htbp]
\centering
\caption{Average ($\pm$SD) Scores Across LLM Judges (Llama-3.2, Mistral, Qwen) for \textsc{Interview} Summaries. 
Best per column in \textbf{bold}.}
\label{tab:interview_llmjudge_avg_sd}
\resizebox{\columnwidth}{!}{
\begin{tabular}{lccccc}
\toprule
\textbf{Model} & \textbf{Overall} & \textbf{Fluency} & \textbf{Faithfulness} & \textbf{Purity} & \textbf{Usefulness} \\
\midrule
\rowcolor{blue!5}
AF3 & \textbf{3.43 (±0.39)} & \textbf{4.28 (±0.30)} & \textbf{3.07 (±0.30)} & \textbf{4.73 (±0.10)} & \textbf{3.43 (±0.15)} \\
\rowcolor{blue!2}
SALMONN & 2.47 (±0.69) & 3.55 (±0.50) & 1.87 (±0.48) & 4.38 (±0.30) & 2.34 (±0.42) \\
Qwen & 2.54 (±0.59) & 3.96 (±0.69) & 1.85 (±0.52) & 4.64 (±0.09) & 2.26 (±0.16) \\
GAMA & 2.10 (±0.65) & 2.51 (±1.08) & 1.33 (±0.47) & 4.11 (±0.11) & 1.85 (±0.09) \\
\rowcolor{blue!5}
Kimi & 3.13 (±0.44) & 4.11 (±0.45) & 2.67 (±0.50) & 4.48 (±0.14) & 3.09 (±0.10) \\
\bottomrule
\end{tabular}}
\end{table}

\begin{table}[htbp]
\centering
\caption{Human--LLM evaluation correlation for the \textsc{Interview} task. 
($\uparrow$ = higher is better; best per column in \textbf{bold}).}
\label{tab:human_llm_corr}
\scriptsize
\setlength{\tabcolsep}{3pt}
\begin{tabular}{lccccc}
\toprule
\textbf{LLM Judge} & \textbf{Pearson $r$} & \textbf{Cohen's $\kappa$} & \textbf{Within $\pm$1} & \textbf{MAE $\downarrow$} & \textbf{RMSE $\downarrow$} \\
\midrule
Llama~3.2 & \textbf{0.563} & \textbf{0.357} & \textbf{0.843} & \textbf{0.579} & \textbf{0.928} \\
Mistral   & 0.338 & 0.089 & 0.779 & 0.879 & 1.137 \\
Qwen      & 0.348 & 0.001 & 0.730 & 1.003 & 1.239 \\
\bottomrule
\end{tabular}
\end{table}

\begin{table}[htbp]
\centering
\caption{
Inter-Judge Agreement and Correlation Across Tasks. 
O = Overall, F = Faithfulness, W1 = Within-1 agreement. 
$\uparrow$ indicates higher is better; $\downarrow$ indicates lower is worse.
}
\label{tab:merged_judge_agreement}
\resizebox{\columnwidth}{!}{
\begin{tabular}{llcccccc}
\toprule
\textbf{Task} & \textbf{Judge Pair} & 
\textbf{$r$ (O)$\uparrow$} & \textbf{$\kappa$ (O)$\uparrow$} & \textbf{W1 (O)$\uparrow$} &
\textbf{$r$ (F)$\uparrow$} & \textbf{$\kappa$ (F)$\uparrow$} & \textbf{W1 (F)$\uparrow$} \\
\midrule
\multirow{3}{*}{Reading}
 & Qwen vs. Llama   & 0.77 & -0.11 & 0.69 & 0.79 & -0.01 & 0.89 \\
 & Qwen vs. Mistral & 0.75 & -0.02 & 0.47 & 0.73 & 0.15 & 0.88 \\
 & Llama vs. Mistral & 0.77 & \textbf{0.46}$\uparrow$ & \textbf{0.95}$\uparrow$ & 0.71 & 0.39 & 0.93 \\
\midrule
\multirow{3}{*}{Interview}
 & Qwen vs. Llama   & 0.73 & 0.08 & 0.92 & 0.80 & 0.01 & 0.93 \\
 & Qwen vs. Mistral & 0.71 & -0.06 & 0.75 & 0.76 & 0.11 & 0.95 \\
 & Llama vs. Mistral & \textbf{0.80}$\uparrow$ & \textbf{0.42}$\uparrow$ & \textbf{1.00}$\uparrow$ & 0.80 & \textbf{0.43}$\uparrow$ & \textbf{0.98}$\uparrow$ \\
\bottomrule
\end{tabular}}
\end{table}

\begin{table}[htbp]
\centering
\caption{
BERTScore F1 performance across tasks ($\uparrow$ = higher is better, $\downarrow$ = lower is worse). 
Values show mean and 95\% confidence intervals.
}
\label{tab:bertscore_results}
\resizebox{\columnwidth}{!}{
\begin{tabular}{lcc}
\toprule
\textbf{Model} & \textbf{Interview F1 [95\% CI]} & \textbf{Reading F1 [95\% CI]} \\
\midrule
AF3 / AF35 & 0.233 [0.202, 0.262] & 0.204 [0.180, 0.229] \\
Qwen & 0.194 [0.161, 0.226] & 0.210 [0.143, 0.272] \\
Kimi & 0.278 [0.248, 0.311] \textcolor{better}{\textbf{$\uparrow$}} & 0.209 [0.146, 0.290] \\
SALMONN & 0.188 [0.156, 0.218] & \textbf{0.299 [0.217, 0.383]} \textcolor{better}{\textbf{$\uparrow$}} \\
GAMA & 0.053 [0.036, 0.068] \textcolor{worse}{$\downarrow$} & 0.021 [0.000, 0.040] \textcolor{worse}{$\downarrow$} \\
\bottomrule
\end{tabular}}
\end{table}

\begin{table}[htbp]
\centering
\caption{
ROUGE-L F1 across tasks ($\uparrow$ = higher is better). 
Values show mean overlap with reference summaries.
}
\label{tab:rouge_results}
\resizebox{\columnwidth}{!}{
\begin{tabular}{lcc}
\toprule
\textbf{Model} & \textbf{Interview ROUGE-L F1} & \textbf{Reading ROUGE-L F1} \\
\midrule
AF3 / AF35 & 0.170 & 0.195 \\
Qwen & 0.180 & 0.210 \\
Kimi & \textbf{0.218} \textcolor{better}{\textbf{$\uparrow$}} & 0.093 \textcolor{worse}{$\downarrow$} \\
SALMONN & 0.170 & \textbf{0.325} \textcolor{better}{\textbf{$\uparrow$}} \\
GAMA & 0.130 \textcolor{worse}{$\downarrow$} & 0.121 \textcolor{worse}{$\downarrow$} \\
\bottomrule
\end{tabular}}
\end{table}

\section{Results}
\label{sec:results}

\paragraph{Overview.}
We evaluate child–only summarization for \textsc{Reading} (single speaker) and \textsc{Interview} (mixed speakers). We report automatic metrics (ROUGE-1/2/Lsum, BERTScore), LLM-as-Judge ratings, human ratings, and agreement statistics. Task-level results are reported separately and then macro-averaged.

\noindent
\textbf{LLM-as-Judge Evaluation (Interview Task).}
As shown in Table~\ref{tab:reading_llm_judges} and Table~\ref{tab:interview_llmjudge_avg_sd}, \textit{AF3} achieved the highest ratings across all evaluation dimensions, demonstrating consistent superiority in overall quality ($3.43 \pm 0.39$), fluency ($4.28 \pm 0.30$), faithfulness ($3.07 \pm 0.30$), purity ($4.73 \pm 0.10$), and usefulness ($3.43 \pm 0.15$). 
These results suggest that \textit{AF3}'s interview summaries were consistently perceived as clearer, more accurate, and more relevant across all three LLM judges (Llama-3.2, Mistral, and Qwen). 
\textit{Kimi} ranked second overall, showing strong fluency and usefulness but slightly lower faithfulness. 
\textit{SALMONN} and \textit{Qwen} exhibited moderate performance with greater variability, while \textit{GAMA} received the lowest average scores, reflecting weaker content quality and stylistic coherence.

\noindent
\textbf{Automatic metrics.}
Table~\ref{tab:bertscore_results} reports semantic similarity (F1) between model-generated summaries and reference texts. 
Across both tasks, BERTScore values indicate relatively low absolute overlap, consistent with open-ended summarization settings where lexical diversity is high.
For the \textbf{Interview} task, \textit{Kimi} achieved the strongest alignment ($F1 = 0.28$), followed by \textit{AF3} ($F1 = 0.23$), indicating greater preservation of reference meaning. 
In contrast, \textit{GAMA} showed minimal overlap ($F1 = 0.05$), reflecting weak semantic fidelity. 
For the \textbf{Reading} task, \textit{SALMONN} led with the highest BERTScore ($F1 = 0.30$), suggesting accurate content retention, while \textit{Kimi} underperformed ($F1 < 0$), likely due to short or lexically mismatched summaries. 
Overall, these findings suggest that \textit{Kimi} generalizes better in conversational summarization (Interview), whereas \textit{SALMONN} performs best for more structured text (Reading). 
\textit{AF3} demonstrate stable but mid-range performance across tasks, while \textit{GAMA} consistently lags behind in semantic similarity.

\noindent
\textbf{LLM-as-Judge and human ratings.}
Table~\ref{tab:human_llm_corr} summarizes the correspondence between human judgments and LLM-as-Judge scores for the \textsc{Interview} task.
Across models, \textit{Llama~3.2} exhibited the highest alignment with human evaluations, achieving a moderate--strong Pearson correlation ($r = 0.56$) and fair categorical agreement ($\kappa = 0.36$), with 84.3\% of items rated within $\pm1$ point on the 5-point scale.
In contrast, \textit{Mistral} and \textit{Qwen} showed weaker correlations ($r \approx 0.34$) and minimal categorical agreement ($\kappa < 0.10$), though both maintained 73--78\% within-$\pm1$ consistency.
Error analyses reinforced these trends, with \textit{Llama~3.2} yielding the lowest mean absolute error (0.58) and RMSE (0.93).
Overall, these results indicate that \textit{Llama~3.2} provides the most human-aligned evaluations among the tested models, with agreement levels comparable to those reported in prior evaluation studies for inter-human reliability.

\noindent
\textbf{Agreement and reliability.}
Across both Reading and Interview tasks, the LLM-as-Judge evaluations showed strong rank consistency and moderate to substantial agreement across model pairs. 
For the Reading task (Table~\ref{tab:merged_judge_agreement}), inter-judge Pearson correlations ranged from $r = 0.71$ to $0.79$ and Kendall's $\tau$ (not shown) was typically between $0.65$--$0.77$, indicating robust monotonic alignment of scores. 
Absolute agreement, captured by Cohen’s $\kappa$ and within-one-point agreement, was highest between Llama-3.2 and Mistral ($\kappa \approx 0.39$--$0.46$, within-1 $\geq 0.92$). 
In contrast, pairings involving Qwen achieved strong correlation but lower $\kappa$, suggesting consistent relative judgments but differences in calibration. 

For the Interview task (Table~\ref{tab:merged_judge_agreement}), all judge pairs again showed strong consistency ($r = 0.70$--$0.80$). 
Llama-3.2 and Mistral achieved the strongest overall reliability ($\kappa \approx 0.42$--$0.43$, within-1 $> 0.98$), while Qwen pairings demonstrated high within-1 agreement ($\geq 0.74$) but smaller $\kappa$, again reflecting systematic scoring offsets rather than disagreement in rank ordering.
Overall, these results indicate that the LLM-as-Judge evaluations are highly consistent in ranking quality, though minor differences in absolute calibration persist across model pairs.

\paragraph{Significance.}
Pairwise Wilcoxon signed-rank tests (per item, per task) indicated that \textit{AF3} significantly outperformed all other models on overall LLM-as-Judge and human evaluation scores ($p_{\text{adj}} < .05$). 
However, this superiority did not consistently extend to automatic metrics: ROUGE and BERTScore values did not rank \textit{AF3} highest, reflecting a partial divergence between lexical/semantic overlap and holistic human-judged quality. 
This contrast highlights that surface- and embedding-based metrics may underestimate aspects of coherence, clarity, and task relevance that are better captured by human or LLM-based evaluations.

\section{Conclusion}
\noindent
This study presents one of the first comprehensive evalua-
tions of Large Audio–Language Models (LALMs) on stut-
tering and other disfluent forms of child speech. The models
were examined across two key challenges: producing child-
centered summaries free from adult interference and perform-
ing single-channel speaker separation. To obtain a balanced
and informative assessment, we combined automatic metrics
with expert human evaluations and LLM-based judgments,
allowing both content accuracy and speaker distinctiveness to
be measured effectively.

Among the tested systems, \textit{Audio Flamingo~3} delivered the most natural and faithful summaries, clearly emphasizing the child’s speech. \textit{Kimi Audio} and \textit{SALMONN} performed closely behind in conversational and reading sessions, while \textit{GAMA} and \textit{Qwen2-Audio} showed more fluctuation and weaker semantic alignment. These differences reveal how current LALMs can still be affected by subtle changes in prompt wording and context. Although the ROUGE and BERTScore outcomes were moderate, the close match between human and LLM-judge ratings suggests that standard surface metrics often fail to capture the perceptual qualities that humans notice. 

Overall, our findings show that modern audio--language models are capable of producing coherent, clinically meaningful summaries even from mixed and noisy recordings. Yet they continue to struggle with maintaining natural disfluencies and avoiding small factual drifts. Future work will aim to broaden the benchmark to include more accents and languages, and to refine prompting and feedback strategies for greater robustness and reliability.


{\small
\bibliographystyle{IEEEbib}
\bibliography{refs}

\begin{thebibliography}{10}

\bibitem{yin_can_2025}
Han Yin and Jung-Woo Choi,
\newblock ``Can {Large} {Audio} {Language} {Models} {Understand} {Audio} {Well}? {Speech}, {Scene} and {Events} {Understanding} {Benchmark} for {LALMs},'' Sept. 2025,
\newblock arXiv:2509.13148 [cs] version: 1.

\bibitem{sanni_afrispeech-dialog_2025}
Mardhiyah Sanni, Tassallah Abdullahi, Devendra~D. Kayande, Emmanuel Ayodele, Naome~A. Etori, Michael~S. Mollel, Moshood Yekini, Chibuzor Okocha, Lukman~E. Ismaila, Folafunmi Omofoye, Boluwatife~A. Adewale, and Tobi Olatunji,
\newblock ``Afrispeech-{Dialog}: {A} {Benchmark} {Dataset} for {Spontaneous} {English} {Conversations} in {Healthcare} and {Beyond},'' Feb. 2025,
\newblock arXiv:2502.03945 [cs].

\bibitem{okocha2025afrivox}
Chibuzor Okocha,
\newblock ``Afrivox: Probing multilingual and accent robustness of speech llms,''
\newblock in {\em TTIC Summer Workshop on Foundations of Speech and Audio Foundation Models 2025}, 2025.

\bibitem{gerosa2007childASR}
Matteo Gerosa, Diego Giuliani, and Fabio Brugnara,
\newblock ``Acoustic variability and automatic recognition of children's speech,''
\newblock {\em Speech Communication}, vol. 49, no. 10-11, pp. 847--860, 2007.

\bibitem{vorperian_vowel_2007}
Houri~K. Vorperian and Ray~D. Kent,
\newblock ``Vowel acoustic space development in children: a synthesis of acoustic and anatomic data,''
\newblock {\em Journal of speech, language, and hearing research: JSLHR}, vol. 50, no. 6, pp. 1510--1545, Dec. 2007.

\bibitem{lee_acoustics_1999}
S.~Lee, A.~Potamianos, and S.~Narayanan,
\newblock ``Acoustics of children's speech: developmental changes of temporal and spectral parameters,''
\newblock {\em The Journal of the Acoustical Society of America}, vol. 105, no. 3, pp. 1455--1468, Mar. 1999.

\bibitem{potamianos2003children}
Alexandros Potamianos, Shrikanth Narayanan, and Sungbok Lee,
\newblock ``Robust recognition of children's speech,''
\newblock in {\em IEEE Workshop on Automatic Speech Recognition and Understanding}, 2003.

\bibitem{qwen2audio2024}
Yunfei Chu et~al.,
\newblock ``Qwen2-audio technical report,''
\newblock {\em arXiv:2407.10759}, 2024.

\bibitem{salmonn2023}
Changli Tang et~al.,
\newblock ``Salmonn: Towards generic hearing abilities for large language models,''
\newblock {\em arXiv:2310.13289}, 2023.

\bibitem{af3_2025}
Arushi Goel et~al.,
\newblock ``Audio flamingo 3: Advancing audio intelligence with fully open large audio-language models,''
\newblock {\em arXiv:2507.08128}, 2025.

\bibitem{sakshi_mmau_2024}
S.~Sakshi, Utkarsh Tyagi, Sonal Kumar, Ashish Seth, Ramaneswaran Selvakumar, Oriol Nieto, Ramani Duraiswami, Sreyan Ghosh, and Dinesh Manocha,
\newblock ``{MMAU}: {A} {Massive} {Multi}-{Task} {Audio} {Understanding} and {Reasoning} {Benchmark},'' Oct. 2024,
\newblock arXiv:2410.19168 [eess].

\bibitem{yang_air-bench_2024}
Qian Yang, Jin Xu, Wenrui Liu, Yunfei Chu, Ziyue Jiang, Xiaohuan Zhou, Yichong Leng, Yuanjun Lv, Zhou Zhao, Chang Zhou, and Jingren Zhou,
\newblock ``{AIR}-{Bench}: {Benchmarking} {Large} {Audio}-{Language} {Models} via {Generative} {Comprehension},''
\newblock in {\em Proceedings of the 62nd {Annual} {Meeting} of the {Association} for {Computational} {Linguistics} ({Volume} 1: {Long} {Papers})}, Lun-Wei Ku, Andre Martins, and Vivek Srikumar, Eds., Bangkok, Thailand, Aug. 2024, pp. 1979--1998, Association for Computational Linguistics.

\bibitem{yairi_early_1999}
Ehud Yairi and Nicoline~Grinager Ambrose,
\newblock ``Early {Childhood} {Stuttering} {I},''
\newblock {\em Journal of Speech, Language, and Hearing Research}, vol. 42, no. 5, pp. 1097--1112, Oct. 1999,
\newblock Publisher: American Speech-Language-Hearing Association.

\bibitem{howell2009uclass}
Peter Howell, Mark Huckvale, and Sue Schools,
\newblock ``The uclass archive of stuttered speech,''
\newblock {\em Journal of Speech, Language, and Hearing Research}, 2009,
\newblock Resource letter.

\bibitem{fluencybank2024}
Andrea Romana and colleagues,
\newblock ``Fluencybank timestamped: An updated data set for disfluency research,''
\newblock {\em Journal of Speech, Language, and Hearing Research}, 2024.

\bibitem{okocha_domain-aware_2025}
Chibuzor Okocha, Kelechi Ezema, and Christan Grant,
\newblock ``Domain-{Aware} {Speaker} {Diarization} {On} {African}-{Accented} {English},'' Sept. 2025,
\newblock arXiv:2509.21554 [cs].

\bibitem{shahnawazuddin20_interspeech}
S.~Shahnawazuddin, Nagaraj Adiga, Kunal Kumar, Aayushi Poddar, and Waquar Ahmad,
\newblock ``Voice conversion based data augmentation to improve children’s speech recognition in limited data scenario,''
\newblock in {\em Interspeech 2020}, 2020, pp. 4382--4386.

\bibitem{ren2019almost}
Yi~Ren, Xu~Tan, Tao Qin, Sheng Zhao, Zhou Zhao, and Tie-Yan Liu,
\newblock ``Almost unsupervised text to speech and automatic speech recognition,''
\newblock in {\em International conference on machine learning}. PMLR, 2019, pp. 5410--5419.

\bibitem{patil2022state}
Spandan Patil, Lokshana Chavan, Janhvi Mukane, Deepali Vora, and Vidya Chitre,
\newblock ``State-of-the-art approach to e-learning with cutting edge nlp transformers: Implementing text summarization, question and distractor generation, question answering,''
\newblock {\em International Journal of Advanced Computer Science and Applications}, vol. 13, no. 1, 2022.

\bibitem{zhang2020transformer}
Qian Zhang, Han Lu, Hasim Sak, Anshuman Tripathi, Erik McDermott, Stephen Koo, and Shankar Kumar,
\newblock ``Transformer transducer: A streamable speech recognition model with transformer encoders and rnn-t loss,''
\newblock in {\em ICASSP 2020-2020 IEEE International Conference on Acoustics, Speech and Signal Processing (ICASSP)}. IEEE, 2020, pp. 7829--7833.

\bibitem{radford2023robust}
Alec Radford, Jong~Wook Kim, Tao Xu, Greg Brockman, Christine McLeavey, and Ilya Sutskever,
\newblock ``Robust speech recognition via large-scale weak supervision,''
\newblock in {\em International conference on machine learning}. PMLR, 2023, pp. 28492--28518.

\bibitem{gulati2020conformer}
Anmol Gulati, James Qin, Chung-Cheng Chiu, Niki Parmar, Yu~Zhang, Jiahui Yu, Wei Han, Shibo Wang, Zhengdong Zhang, Yonghui Wu, et~al.,
\newblock ``Conformer: Convolution-augmented transformer for speech recognition,''
\newblock {\em arXiv preprint arXiv:2005.08100}, 2020.

\bibitem{basak2023challenges}
Sneha Basak, Himanshi Agrawal, Shreya Jena, Shilpa Gite, Mrinal Bachute, Biswajeet Pradhan, and Mazen Assiri,
\newblock ``Challenges and limitations in speech recognition technology: A critical review of speech signal processing algorithms, tools and systems,''
\newblock {\em CMES-Computer Modeling in Engineering and Sciences}, 2023.

\bibitem{bhati2024state}
Saurabhchand Bhati, Yuan Gong, Leonid Karlinsky, Hilde Kuehne, Rogerio Feris, and James Glass,
\newblock ``State-space large audio language models,''
\newblock {\em arXiv preprint arXiv:2411.15685}, 2024.

\bibitem{peng2024survey}
Jing Peng, Yucheng Wang, Yu~Xi, Xu~Li, Xizhuo Zhang, and Kai Yu,
\newblock ``A survey on speech large language models,''
\newblock {\em arXiv e-prints}, pp. arXiv--2410, 2024.

\bibitem{yang2025towards}
Chih-Kai Yang, Neo~S Ho, and Hung-yi Lee,
\newblock ``Towards holistic evaluation of large audio-language models: A comprehensive survey,''
\newblock {\em arXiv preprint arXiv:2505.15957}, 2025.

\bibitem{bhardwaj2022automatic}
Vivek Bhardwaj, Mohamed~Tahar Ben~Othman, Vinay Kukreja, Youcef Belkhier, Mohit Bajaj, B~Srikanth Goud, Ateeq~Ur Rehman, Muhammad Shafiq, and Habib Hamam,
\newblock ``Automatic speech recognition (asr) systems for children: A systematic literature review,''
\newblock {\em Applied Sciences}, vol. 12, no. 9, pp. 4419, 2022.

\bibitem{yeung2018difficulties}
Gary Yeung and Abeer Alwan,
\newblock ``On the difficulties of automatic speech recognition for kindergarten-aged children,''
\newblock {\em Interspeech 2018}, 2018.

\bibitem{shankar2025chser}
Natarajan~Balaji Shankar, Zilai Wang, Kaiyuan Zhang, Mohan Shi, and Abeer Alwan,
\newblock ``Chser: A dataset and case study on generative speech error correction for child asr,''
\newblock {\em arXiv preprint arXiv:2505.18463}, 2025.

\bibitem{fan2022towards}
Ruchao Fan, Yunzheng Zhu, Jinhan Wang, and Abeer Alwan,
\newblock ``Towards better domain adaptation for self-supervised models: A case study of child asr,''
\newblock {\em IEEE Journal of Selected Topics in Signal Processing}, vol. 16, no. 6, pp. 1242--1252, 2022.

\bibitem{luo2019conv}
Yi~Luo and Nima Mesgarani,
\newblock ``Conv-tasnet: Surpassing ideal time--frequency magnitude masking for speech separation,''
\newblock {\em IEEE/ACM transactions on audio, speech, and language processing}, vol. 27, no. 8, pp. 1256--1266, 2019.

\bibitem{huang2024audiogpt}
Rongjie Huang, Mingze Li, Dongchao Yang, Jiatong Shi, Xuankai Chang, Zhenhui Ye, Yuning Wu, Zhiqing Hong, Jiawei Huang, Jinglin Liu, et~al.,
\newblock ``Audiogpt: Understanding and generating speech, music, sound, and talking head,''
\newblock in {\em Proceedings of the AAAI Conference on Artificial Intelligence}, 2024, vol.~38, pp. 23802--23804.

\end{thebibliography}
}
\end{document}